\documentclass{mpe_report}
\usepackage{psfig}

\begin{document}
\title{The X-ray luminosity function of galaxies and its evolution}
\author{I. Georgantopoulos\inst{1} \and P. Tzanavaris\inst{1}} 
\institute{National Observatory of Athens, I. Metaxa 1, 15236 Penteli, Greece}  
\maketitle

\begin{abstract}
We compile one of the largest ever samples to probe 
 the X-ray normal galaxy luminosity function and its evolution 
 with cosmic time. In particular, we select 207 galaxies (106 late and 
 101 early-type systems)
 from the Chandra Deep Field North and South surveys, the Extended Chandra Deep 
 Field South and the XBOOTES survey. We derive the luminosity function 
 separately for the total (early+late), the early and the late-type samples 
 using both a parametric maximum likelihood method, and a variant of the 
 non-parametric $1/V_m$ method.  Although the statistics is limited, 
 we find that the total (early+late) galaxy sample 
 is consistent with a Pure Luminosity evolution model where the luminosity 
 evolves according to $L(z) \propto  (1+z)^{2.2}$. The late-type systems 
 appear to drive this trend while the early-type systems 
 show much weaker evidence for evolution. 
 We argue that the X-ray evolution of late-type systems is consistent 
 with that of blue galaxies in the optical. In contrast 
 there is a mismatch between the X-ray evolution of 
 early-type systems and that of red galaxies at optical wavelengths.  
\end{abstract}

\section{Introduction}

 The Chandra and XMM missions have led to great progress 
 in our understanding of X-ray emission processes in 
 normal galaxies (see review by Fabbiano, 2005).
 The X-ray emission in the most luminous early-type systems 
 comes mainly from hot gas, while in the less luminous 
 ones, Low-Mass X-ray binaries (LMXRB) prevail. 
 On the other hand, in late-type systems the X-ray binaries dominate 
 the X-ray emission. In the most actively star-forming ones, 
 it is the High-Mass X-ray binaries (HMXRB) in particular which 
 dominate the X-ray energy budget. 
 In star-forming galaxies, the total galaxy X-ray luminosity 
 is an excellent indicator of star-formation (e.g. Ranalli et al. 2003, 
 Gilfanov et al.  2004).  
 
 The study of the galaxy luminosity function at X-ray wavelengths provides the 
 opportunity to study, as an ensemble, the evolution of these systems 
 over cosmological timescales. For example, it is important to 
 investigate whether the evolution of the X-ray emission for 
 star-forming systems is the same as in optical and infrared wavelengths. 
 Ghosh \& White (2001) suggest that there may be a time-lag because 
 of the onset of the LMXRB, a few Gyr after the star-burst event.
 In this respect the X-ray luminosity function has important
 implications for the evolution of X-ray binaries.    

 One predicament, in the construction of X-ray selected galaxy samples is that, 
  galaxies are intrinsically X-ray faint sources. 
 This implies that we need to probe either very deep in flux or
 over a wide field-of-view to gather sufficiently large samples.  
 The deep Chandra 
 observations in the area of the Hubble Deep Field North 
 provided the first X-ray selected galaxy sample (Hornschemeier et al. 2003).  
 Norman et al. (2004) derived the X-ray luminosity function finding 
 hints for pure luminosity evolution. The study of Norman et al. (2004) 
 pertains to the total galaxy sample i.e. including both early and late type. 
 At the same time XMM surveys   
 with their large field-of-view,  have helped to constrain the 
 galaxy luminosity function in the local Universe at a median redshift
 of $z\sim 0.1$ (Georgakakis et al. 2004, Georgantopoulos et al. 2005, 
 Georgakakis et al. 2006). The luminosity function 
 is well represented by either a Schechter or a log-norm form.  

 Here, we present the galaxy luminosity function 
 derived from the analysis of  a large Chandra sample of galaxies     
 consisting both of Deep Fields and wide area surveys.
 $\rm H_0=72 \ km~s^{-1} \ Mpc^{-1}, \Omega_m=1/3$ and $\rm \Omega_\Lambda=2/3$ are 
 adopted.     

 \section{Data and Method}
 \subsection{The Data}
 We compile our sample from the Chandra Deep Field North and 
 South surveys, CDF-N and CDF-S,
 the Extended Chandra Deep Field South (ECDF-S) and the XBOOTES survey. 
 The 2Ms exposure of the CDF-N (Alexander et al. 2003) represents 
 the deepest observations
 of the X-ray Universe ever, reaching a flux depth of $\sim 2 \times 10^{-17}$
 $\rm erg~cm^{-2}~s^{-1}$ 
 in the 0.5-2 keV band but only over a small area ($\rm \sim 400 \ arcmin^{2}$).
  The CDF-S (Giaconni et al. 2002) reaches a flux limit of 
 $4 \times 10^{-17}$ $\rm erg~cm^{-2}~s^{-1}$
 while the ECDF-S (Lehmer et al. 2006) covers four ACIS-S fields at a flux limit 
 about one order of magnitude brighter than the CDF-N. 
 Finally, the XBOOTES survey (Kenter et al. 2005) 
 covers an area of about 10 $\rm deg^{2}$ but reaching 
 a flux limit about 200 times brighter than the CDF-N. 
 The optical photometric data for the CDF-N, CDF-S, ECDF-S and XBOOTES data 
 come from Capak et al. (2004), Zheng et al. (2004), COMBO-17 (Wolf et al. 2003)
 and the SDSS survey respectively. 

\subsection{Selection Criteria}  
 We use sources detected in the soft 0.5-2 keV band 
 as galaxies are preferentially soft X-ray emitters (e.g. Levenson et al. 2001). 
 We select galaxies using the following criteria:
 \begin{itemize}
 \item{Low X-ray to opical flux ratio, $f_x/f_R<-1$. We choose 
 this limit in order to minimize any possible loss of early-type galaxies 
 at high redshift. Ptak et al. (2007) have shown that the 
 K-correction results in an increase of the $f_x/f_R$ ratio of early-type systems 
 at high redshift;}
 \item{Soft X-ray Spectrum. We choose the hardness ratio, HR=(H-S)/(H+S),
  where H and S are the count rates in the hard and soft bands respectively,
 to correspond to a spectrum with a photon index of $\Gamma>1.4$; }
 \item{Low X-ray luminosity, $\rm L(0.5-2 keV)<10^{42}$ $\rm erg~s^{-1}$;
 but see Tzanavaris et al. 2006}
 \item{Extended optical images. All star-like sources have been 
 discarded.}
 \end{itemize} 
 We note that our criteria are very similar to those applied by 
 Bauer et al. (2005) in their \lq pessimistic\rq\ galaxy sample.

\subsection{The Sample}
 We end up with 207 sources classified as galaxies.
The combination of the four surveys provides excellent coverage of 
 the luminosity-redshift space (see Fig. 1), with the XBOOTES survey
 detecting luminous systems at nearby redshifts, whilst the 
 Chandra deep fields yield less luminous systems at higher redshift. 
 We classify our galaxies as early-type or late-type  
by means of the broad-band colours. We use 
 template spectral energy distributions in 
 {\sl HYPERZ} (Bolzonella et al. 2000). 
 {\sl HYPERZ} performs a $\chi^2$ minimization 
 to fit a template spectral energy distribution. 
We find 101 and 106 early and late-type galaxies, 
 respectively.  
 
\subsection{AGN contamination}
One of the most important aspects in the compilation of 
 an X-ray galaxy sample is the possible 
 contamination of the X-ray light from an AGN.
 Georgakakis et al. (2007) have explored in depth 
 this problem. They suggest that the best approach,
 in the case of late-type systems, is to use mid-IR data. 
 Then the AGN can be easily found 
 as they present excess X-ray emission over their mid-IR emission.
 Unfortunately, we do not have mid-IR data available for all 
 our fields. Nevertheless, from the CDF-N field, where 
 Spitzer data are available, we can derive 
 a rough estimate of the expected contamination.
We find that the expected contamination is lower than 20\%,
 for the selection criteria we used above i.e. $L_x<10^{42}$ $\rm erg~s^{-1}$,
 $\rm f_x/f_R<-1$, $HR<0$.

\begin{figure}
\centerline{\psfig{file=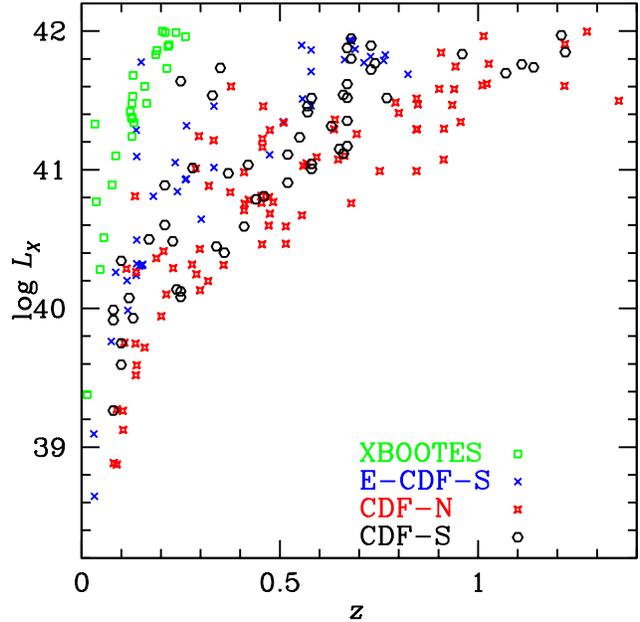,width=8.8cm,clip=} }
\caption{The luminosity-redshift coverage.
\label{LZ}}
\end{figure}

\section{The Luminosity Function} 

\subsection{Derivation} 
 We use the non-parametric method of Page \& Carrera (2000), 
 which is a variant of the classical $1/V_{m}$ method (Schmidt 1968),
 to produce the binned luminosity function. 
 We also derive the luminosity function using the 
 parametric maximum-likelihood method (Tammann et al. 1979).
 Although the maximum-likelihood method is usually prefered 
 as it does not bin the data,  
 it presents the disadvantage that the model has to be
 chosen a priori. Furthermore this method does not 
 provide a goodness of fit.  
 We fit a Schechter form to the luminosity function
 which is known to parameterize very well 
 the optical luminosity function  
 
 \begin{equation}
 \phi(L)= \phi_\star (\frac{L}{L_\star})^{-\alpha}~exp(-\frac{L}{L_\star})~ d(\frac{L}{L_\star})  \ ,
 \end{equation} 

 where $\alpha$ denotes the slope at faint luminosities. At bright luminosities 
 $\rm L> L_\star$  the luminosity function drops exponentially. 
 We assume that the characteristic luminosity $\rm L_\star$ evolves  
 according to a Pure Luminosity Evolution model: 
 $L_\star(z) = L_\star(0)~(1+z)^k$ where the evolution index 
 is $k=0$ in the case of no evolution.     

 \subsection{The total sample}
 We first apply the above methods to the total sample, dividing  
 into three redshift bins (see Fig. 2). The median redshifts for the 
 three bins are 0.13, 0.38 and 0.78. 
 The binned luminosity function 
 shows a clear hint for evolution with redshift.
 The maximum likelihood method confirms this result. 
 The free parameters are $L\star$, $\alpha$ and the evolution index $k$.  
 The evolution index derived is $k= 2.2\pm 0.3$ (see table 1).  

\begin{figure}
\centerline{\psfig{file=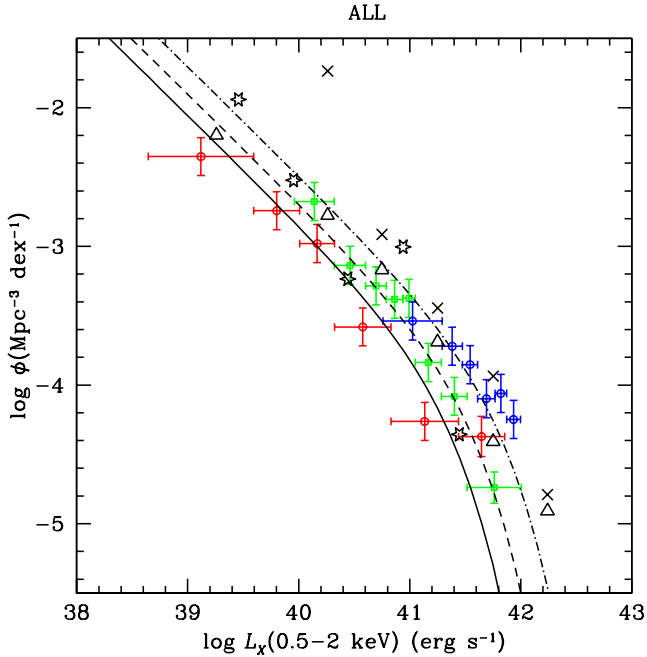,width=8.8cm,clip=} }
\caption{The galaxy luminosity function for the 
 total sample (early$+$late type) in three redshift bins:
 red (z$<$0.2), green ($0.2<z<0.6$), blue ($0.6<z<1.4$).
 The lines represent the maximum likelihood fits 
 to the above redshift intervals.  
 The stars denote the results of Kim et al. (2006); the 
 crosses and triangles represent the results of 
 Norman et al. (2004).  
\label{total}}
\end{figure}

\begin{table*}
\caption{ML fit results for a Schechter function} 
\begin{tabular}{ccccc}
         \hline
            Sample    &  $\log L_\star$  & $\alpha$      &  $k$               & $\phi_\star$ \\
                      &                 &               &    &$\ln (10) \times 10^{-4}$ Mpc$^{-3}$ dex$^{-1}$\\
         \hline
             Total    & $41.24\pm0.02$ & $-1.79\pm0.06$ & $2.2\pm 0.3$         & 1.24        \\
             Early    & 41.68          & $-1.7\pm0.1$   &  $0.0^{+0.8}_{-0.7}$ & 0.45        \\ 
             Late     & 41.23          & $-2.0\pm0.1$   & $2.5^{+1.1}_{-1.8}$  & 0.39        \\
         \hline
   \end{tabular}
\end{table*}

\subsection{The early and late-type samples}
Here, we derive the luminosity function for the early and late-type 
 samples. 
 It is important to study the evolution separately for the 
 two subsamples, as these may evolve in a different manner, exactly 
 as it is the case in optical wavelengths. 
We divide into two redshift bins only, owing to small 
 number statistics. The binned luminosity functions are shown in Fig. 3 
 both for early- and late-type galaxies.
 There is evidence for evolution in the late-type sample between 
 the median redshifts of the two bins z=0.14 and z=0.67. 
 In contrast, there is no clear evidence for evolution in the case of 
 early-type galaxies. 
 Admittedly, from the inspection of the binned luminosity function, 
 it is very difficult to tell whether the evolution can be described 
 by Pure Luminosity evolution alone. Some density evolution especially
 at bright luminosities may also be taking place. 
    
The maximum-likelihood results (assuming 
 Pure Luminosity evolution) are given in 
 table 1. Note that because of small number
 statistics and the large number of free parameters, 
 the characteristic luminosity $\rm L_\star$ was fixed to the value
 derived in the local luminosity function of Georgantopoulos et al. (2005).

 It is interesting to compare our results with the 
 X-ray luminosity function  
 predicted from the optical data (solid line in Fig. 3). 
 This {\it predicted} X-ray luminosity function can be 
 estimated as the convolution of the optical luminosity 
 function and the $L_x/L_o$ relation for the corresponding 
 sample of galaxies (e.g. Georgantopoulos et al. 1999). 
 We use the optical luminosity functions from the SDSS 
 (Nakamura et al. 2004) and the $L_x/L_o$ relations in 
 Shapley et al. (2001). 
 It appears that the optical luminosity functions do not provide 
 a good match to the actual X-ray data, at least in the case 
 of the early-type sample. 
 The optical luminosity function gives an excess of 
 low luminosity late-type objects. On the other hand the 
 optical luminosity function predicts a larger density 
 of early-type galaxies 
 than actually observed at all luminosities. 

\begin{figure}
\centerline{\psfig{file=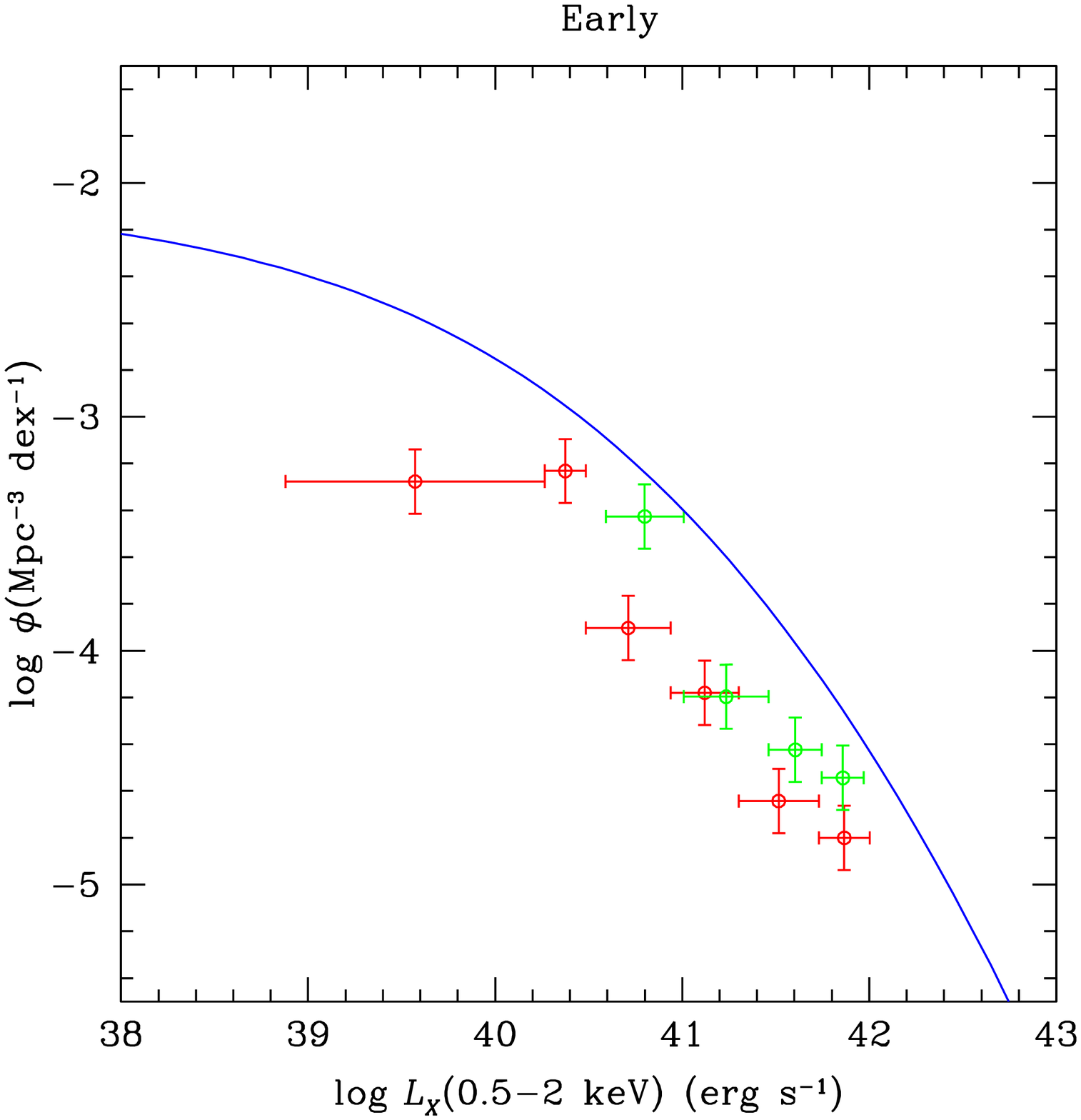,width=8.8cm,clip=} }
\centerline{\psfig{file=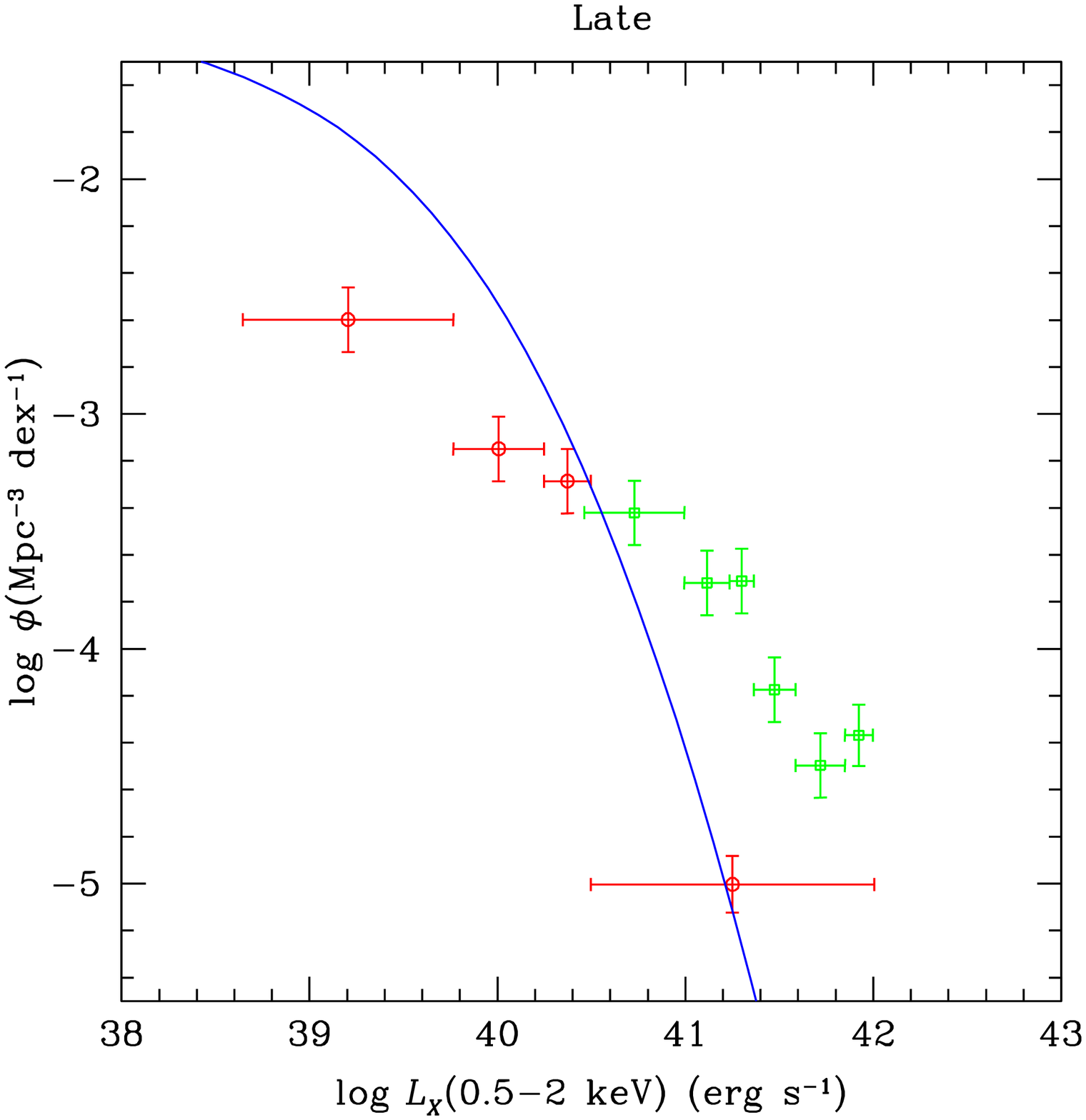,width=8.8cm,clip=} }
\caption{Upper panel: The early-type binned luminosity function 
 in two redshift bins: red $z<0.4$, green $0.4<z<1.4$. 
 Lower panel: The late-type binned luminosity function 
 for the same redshifts bins as above. The blue line 
 denotes the X-ray luminosity function as predicted from the 
 convolution of the optical SDSS luminosity function and the $L_x/L_o$
 relation (see text).     
\label{early-late}}
\end{figure}

\section{Discussion}

\subsection{The cosmic evolution} 
 The X-ray luminosity function for galaxies has been derived 
 for the first time by Norman et al. (2004). They 
 used data from the Chandra Deep Fields (CDF-N and CDF-S)  but without 
 discriminating for early and late type systems. Norman et al. (2004) 
 find strong pure luminosity evolution with an evolution 
 index $k=2.7$. Recently, Ptak et al. (2007) constructed again 
 the luminosity function in the same fields. 
 Their more conservative selection criteria result in 
 a smaller sample compared to that of Norman et al. (2004).  
 They derive the non-parametric luminosity function separately 
 in various redshift bins, estimating the evolution index $k$ 
 from the difference in the $L_\star$. Their late-type sample shows 
 strong Pure Luminosity Evolution,
 with $k=2.3^{+0.8}_{-0.8}$. In the case of the early-type sample 
 they find some mild evolution $k=1.6^{+1.1}_{-1.0}$, which 
 however is not inconsistent with our results.  
 Georgakakis et al. (2006) derived the evolution 
 in the Chandra Deep Fields using number counts, again
 separately for early- and late-type samples. 
 They found evolution only for late-type systems ($k=2.7\pm0.35$) while
 the early-type ones are consistent with no evolution ($k=0.6^{+0.8}_{-0.6}$).
 Georgakakis et al. (2007) applying a very careful  
 galaxy selection, aiming to minimize AGN contamination, 
 find an evolution index of $k=2.4$ for the star-forming 
 galaxies in the CDF-N. From the above, it appears then that the 
 results are converging
 as far as the evolution is concerned: under the 
 assumption of luminosity evolution the late-type systems 
 evolve strongly, while the early-type ones show milder 
 or possibly no evolution.    
 However, clearly better statistics is needed to test whether 
 more complex forms of evolution take place, as for example
 Luminosity Dependent Density Evolution.

\subsection{Interpretation and Comparison with the optical}

The optical luminosity function has been derived by Faber et al. (2007)
 for blue and red galaxies using DEEP2 data. They find that both 
 blue systems evolve vigorously over cosmic time
 between a redshift of z=1 and today, following pure luminosity evolution 
 with $k\sim2.7$. The red galaxies evolve in a different manner. 
 Although their luminosity function evolves in luminosity 
 in a similar fashion,its normalization is higher at redshift z=0. 
 Now assuming  that roughly the blue galaxies correspond
 to our late-type systems and the red to the early-type ones, 
 we can make a comparison between the galaxy evolution
 probed in X-ray and optical wavelengths. 
 The late-type galaxies have a very similar evolution in both 
 the optical and the X-ray regime. This suggests that the 
 X-ray emission in these galaxies is dominated by HMXRB.
 Indeed HMXRB evolve on Myr scale and therefore they trace 
 closely the bulk of the optical and IR emission coming 
 from massive stars.   
 In contrast, there is a mismatch in the evolution 
 of early-type systems witnessed in optical 
 and X-ray wavelengths. The absence of vigorous 
 evolution in X-ray wavelengths between 
 z=0.7 and today ($\sim 6$ Gyr), is much longer   
 than the expected lifetimes of LMXRBs ($0.1-1$Gyr),
 placing interesting constraints on the formation 
 of these systems.

%
%
   

\end{document}